\documentclass[aps,prb,10pt,twocolumn,floatfix,showpacs,superscriptaddress,longbibliography]{revtex4-1}

\usepackage{graphicx,amsmath,amsfonts,bm}
\usepackage[colorlinks=true, citecolor=blue, linkcolor=blue, urlcolor=blue]{hyperref}
\usepackage[all]{hypcap}

\begin{document}

\title{Conductance signatures of odd-frequency superconductivity in quantum spin Hall systems using a quantum point contact}
\author{C. Fleckenstein}
\email{christoph.fleckenstein@physik.uni-wuerzburg.de}
\affiliation{Institute of Theoretical Physics and Astrophysics, University of W\"urzburg, 97074 W\"urzburg, Germany}
\author{N. Traverso Ziani}
\affiliation{Institute of Theoretical Physics and Astrophysics, University of W\"urzburg, 97074 W\"urzburg, Germany}
\author{B. Trauzettel}
\affiliation{Institute of Theoretical Physics and Astrophysics, University of W\"urzburg, 97074 W\"urzburg, Germany}
\begin{abstract}
Topological superconductors give rise to unconventional superconductivity, which is mainly characterized by the symmetry of the superconducting pairing amplitude. However, since the symmetry of the superconducting pairing amplitude is not directly observable, its experimental identification is rather difficult. In our work, we propose a system, composed of a quantum point contact and proximity induced $s$-wave superconductivity at the helical edge of a two dimensional topological insulator, for which we demonstrate the presence of odd-frequency pairing and its intimate connection to unambiguous transport signatures. Notably, our proposal requires no time-reversal symmetry breaking terms. We discover the domination of crossed Andreev reflection over electron cotunneling in a wide range of parameter space, which is a quite unusual transport regime.
\end{abstract}

\maketitle

\section{Introduction}
Two dimensional topological insulators (TIs), due to strong spin orbit-coupling, provide surface states with perfect spin-momentum locking \cite{KaneMele2005, KaneMele2005b, Bernevig2006}. Over the last decade of research, convincing evidence has been reported, proving the existence of those topological states in HgCdTe/HgTe \cite{Konig, Roth2009} and InAs/GaSb quantum wells \cite{CLiu2008, Knez2011, Knez2012, Knez2015, VSPribbiag}. As long as electron-electron interactions are weak \cite{CWu2006, Geissler2014, NTZiani2015, NTZiani2016}, the presence of time-reversal (TR) symmetry inevitably leads to a suppression of backscattering processes and thus provides dissipationless transport. This feature is usually attributed to superconductivity (SC). Interestingly, it is possible to combine both effects by proximity-induce $s$-wave superconductivity into the helical edge of a quantum spin Hall insulator (QSHI) \cite{Knez2012,  VSPribbiag, LFu2008, LFu2009, SHart}. There, the combination of conventional superconducting order and spin-momentum locking gives rise to unconventional superconductivity. 

The associated order parameter is the superconducting pairing amplitude $\mathcal{F}$, which is directly related to the anomalous part of the retarded Green function. According to the classification pioneered by Berezinskii \cite{Berezinskii}, the pairing amplitude $\mathcal{F}$ has to be totally antisymmetric under the exchange of all quantum numbers of the two constitutent fermionic field operators. Having spin, orbit and frequency as characteristics, this yields a set of four symmetry classes \cite{YTanaka2007, YTanaka2012, YTanaka2007a, GerdSchon, YTanaka2007b}. While conventional BCS superconductors are even in frequency, 
it has been shown that a special kind of unconventional superconductivity, that is odd in frequency, arises quite ubiquitously in spatially non-uniform systems in which spin rotation invariance is broken \cite{YTanaka2012, Linder2017}, such as SC-TI heterojunctions \cite{JCayao}, or heterojunctions including additional ferromagnetic ordering \cite{AMBlack, YAsano2013, Crepin2015, Keidel2018}. However, since the symmetry of the pairing amplitude is not a quantum mechanical observable, it is challenging to unambigiously probe odd-frequency pairing.

The first experimental signature attributed to odd-frequency SC has been identified as a long range proximity effect in ferromagnetic Josephson junctions \cite{FSBergeret2005, Alidoust2010, TSKhaire2011}. Subsequently, a paramagnetic Mei{\ss}ner effect has been proposed if odd-frequency SC is present \cite{Asano2010a, Alidoust2014, Bernardo2015}. Recently, it has furthermore been demonstrated that odd-frequency SC can be directly assigned to particular transport properties, using a QSHI in proximity to both, $s$-wave SC and ferromagnetic ordering \cite{Crepin2015, Alidoust2017}. At the helical edge, the ferromagnetic ordering allows for odd-frequency equal-spin pairing, intimately related to crossed Andreev reflection (CAR), where an incident electron is transmitted through the scattering region as a hole and scattered into a second, spatially separated lead \cite{Beckmann2004, SRusso, Morten2006, JCassyol, PBurset2016}. Spin-momentum locking then implies the equivalence of this transport channel with the creation of an equal-spin triplet Cooper pair in the heterostructure \cite{CBen2006, BHWu2014}.
Using resonances between hybridizing Majorana bound states, it has been proposed that CAR can overcome electron cotunneling (EC) across the junction, and thus yields a smoking-gun evidence of odd-frequency SC. However, despite interesting applications \cite{Qi2008, Jukka2011, Orth2015, CFleckenstein2016, SPorta2018}, it is experimentally challenging to induce ferromagnetic ordering in 2D TIs and more feasible setups are needed, in which TR symmetry is preserved.

\begin{figure}
\centering
\includegraphics[scale=0.55]{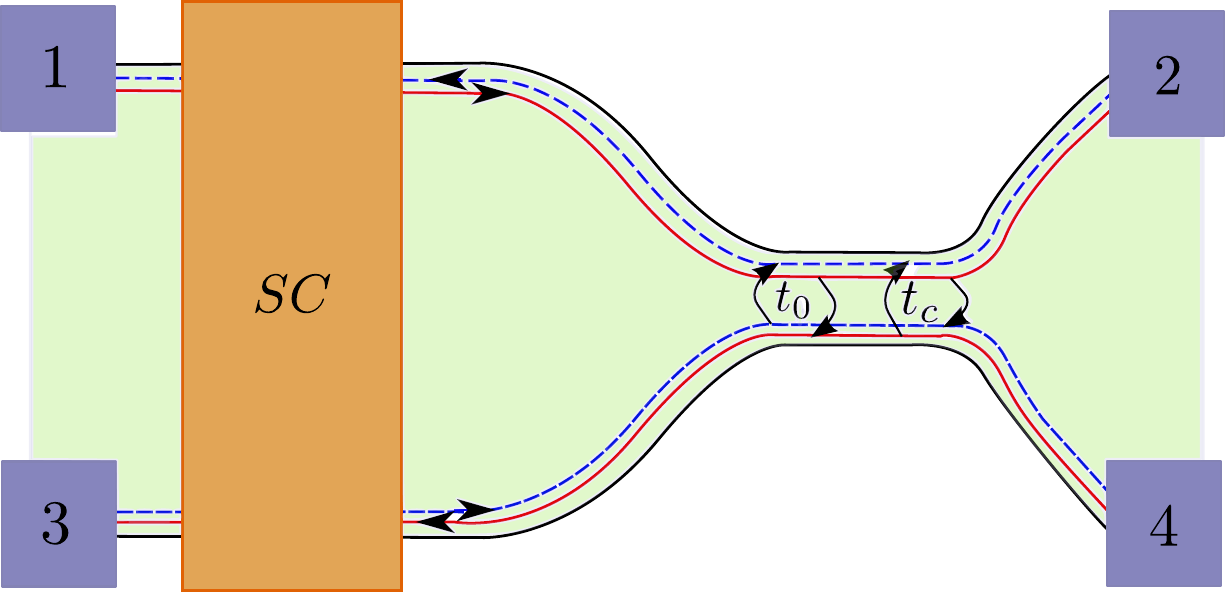}
\caption{Schematic of the system: Each edge of a quantum spin Hall sample is coupled to separate contacts 1-4. The sample itself contains a heterostructure, composed of a region with proximity induced $s$-wave superconductivity and a quantum point contact including the terms $t_0$ and $t_c$, defined in Eqs. (\ref{Eq:QPC1}) and (\ref{Eq:QPC2}).}
\label{Fig:sys}
\end{figure}
In this article, we present a simple system, composed of proximity induced s-wave pairing and a quantum point contact (QPC) at the helical edge of a 2D TI (see Fig. \ref{Fig:sys}). This setup has recently been investigated with an emphasis on Kramers pairs of Majorana fermions, present when a superconducing phase shift of $\pi$ is applied between separate superconductors, covering the two edges \cite{Lutchyn2016}. For our purposes, this setup has the crucial advantage that no ferromagnets are involved, but it still contains the desired feature: odd-frequency equal-spin pairing with direct connection to the differential conductance of the system. 
More specifically, we demonstrate that such a QPC can generate equal-spin triplet pairing if axial spin symmetry is broken. Locally, equal-spin correlations are suppressed. However, non-local correlations appear across the QPC/junction. They inevitably lead to non-vanishing CAR between contacts 1 and 2 in Fig. \ref{Fig:sys}. Additionally, the QPC provides three accessible channels for EC. Therefore, the EC per channel is reduced.
The combination of both effects leads to a domination of CAR over EC for a wide range in parameter space. This dominance is observable in the linear (non-local) conductance, as it directly probes the difference between CAR and EC, which hence implies the presence of odd-frequency SC.

The article is structured as follows: In Sec. II, we introduce the model and give a theoretical description of the symmetry of the superconducting pairing amplitude in terms of the corresponding Green function. In Sec III, we analyse the different pairing amplitudes present in our system, while in Sec. IV we discuss their relation to transport channels by means of the differential (non-local) conductance. Finally, we conclude in Sec. V. Technical details are provided in two Appendicies.
\section{Model and Green function}
\subsection{Model}
The system we investigate is formed by the helical edge of a 2D TI, partially covered by an s-wave superconductor and coupled to a QPC, as schematically shown in Fig. \ref{Fig:sys}. The physics of each edge of the 2D TI channel is captured by the effective edge state Hamiltonian
\begin{eqnarray}
H_{\lambda,0}&=&\int dx\Psi_{\lambda}^{\dagger}(x)\mathcal{H}_{\lambda}(x)\Psi_{\lambda}(x)
\end{eqnarray}
with Hamiltonian density
\begin{equation}
\mathcal{H}_{\lambda}(x)=-\lambda \hbar v_F i \partial_x \tau_z\sigma_z-\mu(x) \tau_z\sigma_0+\Delta(x)\tau_x\sigma_0
\end{equation}
and basis
\begin{equation}
\label{Eq:Basis}
\Psi_{\lambda}(x)=(\psi_{\lambda,\uparrow}(x),\psi_{\lambda,\downarrow}(x),\psi_{\lambda,\downarrow}^{\dagger}(x),-\psi_{\lambda,\uparrow}^{\dagger}(x))^T,
\end{equation} 
where $\lambda\in\{+,-\}$ (for upper and lower edge). The Pauli matrices $\tau_{i},~\sigma_{i}$ with $i\in\{x,y,z\}$ act on particle-hole and spin space, respectively. We assume a spatially varying chemical potential $\mu(x)$. Furthermore, we include a proximity induced s-wave pairing by $\Delta(x)$ with the characteristic length scale $\xi_{\Delta}=\hbar v_F/\Delta$. In the following we fix $\hbar v_F=1$.
In the constriction of the quantum point contact, we additionally add two TR invariant scattering processes across the edges \cite{CYJeffrey2009, JCBudich, Dolcini2011, Dolcetto2014, Lutchyn2016}, that are: (i) spin-conserving backscattering 
\begin{equation}
\label{Eq:QPC1}
H_{t_0}\!=\!\int\! dxt_0(x) \!\left(\!\psi^{\dagger}_{+,\uparrow}(x)\psi_{-,\uparrow}(x)+\psi^{\dagger}_{+,\downarrow}(x)\psi_{-,\downarrow}(x)\!\right)+\mathrm{H.c.},
\end{equation}
and (ii) forward scattering, breaking axial spin symmetry \cite{CWu2006, Schmidt2012, Jukka2013, Kainaris2014, Matteo2, Matteo1}
\begin{equation}
\label{Eq:QPC2}
H_{t_c}\!=\!\int\! dxt_c(x)\! \left(\!\psi^{\dagger}_{+,\uparrow}(x)\psi_{-\downarrow}(x)-\psi^{\dagger}_{+,\downarrow}(x)\psi_{-,\uparrow}(x)\!\right)+\mathrm{H.c.}.
\end{equation}
The full Hamiltonian of the system then reads
\begin{equation}
H=\sum_\lambda H_{\lambda,0}+H_{t_0}+H_{t_c}.
\end{equation}
In compact form, we can state the full Hamiltonian density as
\begin{eqnarray}
\label{Eq:HamDens}
\mathcal{H}(x)&=&-i\partial_x s_z\tau_z\sigma_z-\mu(x) s_0\tau_z\sigma_0+\Delta(x)s_0\tau_x\sigma_0\nonumber\\
&+&t_0(x)s_x\tau_z\sigma_0-t_c(x)s_y\tau_z\sigma_y
\end{eqnarray}
using the basis $\Phi(x)=\left(\Psi_+(x),\Psi_-(x)\right)^T$ and the Pauli matrices $s_i$ with $i\in\{x,y,z\}$ acting on edge space. We model the heterostructure shown in Fig. \ref{Fig:sys} by using piecewise constant potentials $\Delta(x)=\Delta \theta(x)\theta(l_s-x)$, $t_0(x)=t_0\theta(x-x_q)\theta(x_q+l_q-x)$ and  $t_c(x)=t_c\theta(x-x_q)\theta(x_q+l_q-x)$ with $\theta(x)$ the Heaviside function. Here, $\Delta$, $t_0$ and $t_c$ are real and positive parameters, while $x_q$ marks the beginning of the quantum point contact. Furthermore, $l_q$ and $l_s$ are the length of the constriction and the superconductor, respectively.

We can solve the first order differential Schr\"odinger equation $\mathcal{H}(x)\Phi(x)=\omega\Phi(x)$ by integration. From
\begin{eqnarray}
\partial_x \Phi(x)&=&i\big[ h(x)+ s_z\tau_z\sigma_z \omega\big]\Phi(x)
\end{eqnarray}
with
\begin{eqnarray}
h(x)&=&s_z\tau_z\sigma_z\big[\mu(x) s_0\tau_z\sigma_0-\Delta(x)s_0\tau_x\sigma_0\nonumber\\
&-&t_0(x)s_x\tau_z\sigma_0+t_c(x)s_y\tau_z\sigma_y\big],
\end{eqnarray}
we find the general solution
\begin{equation}
\label{Eq:sol1}
\Phi(x)=\mathcal{S}_{\leftarrow}U(x,x_0)\Phi_0(x_0),
\end{equation}
where 
\begin{equation}
\label{Eq:sol2}
U(x,x_0)=\exp\left[i\int_{x_0}^x dx \left(h(x)+s_z\tau_z\sigma_z\omega\right) \right].
\end{equation}
In Eq. (\ref{Eq:sol1}), $\mathcal{S}_{\leftarrow}$ is a spatial-ordering operator, required to order all operators, acting  on $\Phi_0(x_0)$, with their spatial coordinates increasing from right to left \cite{CTimm2011}. As we only apply piecewise constant potentials, we can neglect $\mathcal{S}_{\leftarrow}$ whenever the integration runs within a homogenous section. Together with the condition of continuity of the wavefunction at each interface between sections of different potentials, the scattering problem, defined by the Hamiltonian (\ref{Eq:HamDens}), can be formulated as
\begin{equation}
\label{Eq:scatter}
\Phi_{\mathrm{out},l}(x_q+l_q)=U_{t}(x_q+l_q,x_q)U_{0}(x_q,l_s)U_{sc}(l_s,0)\Phi_{\mathrm{in},l}(0)
\end{equation}
with the propagators $U_t(x_t,x_t')$, $U_0(x_0,x_0')$ and $U_{sc}(x_{sc},x_{sc}')$ defined according to Eq. (\ref{Eq:sol2}) in the bounds $\{x_t,x_t'\}\in[x_q,x_q+l_q]$, $\{x_0,x_0'\}\in[l_s,x_q]$, and $\{x_{sc},x_{sc}'\}\in[0,l_s]$.
The form of the vectors $\Phi_{\mathrm{in},l}(x)$ and $\Phi_{\mathrm{out},l}(x)$ are fixed by spin momentum locking, together with the basis of Eq. (\ref{Eq:Basis}). 

However, we still need to select a channel for an incident particle, whose amplitude we fix to unity \cite{Akhmerov2011}. This procedure allows us to derive eight independent scattering states denoted by the index $l\in[1,8]$. They are classified according to the incoming amplitude: incoming electron/hole from the right/left in edge $+$/$-$. A detailed derivation thereof is provided in App. A. Consequently, an incoming partcile $\chi\in\{e,h\}$ from edge $\lambda$ can be reflected as an electron with amplitude $r_{\chi e}^{\lambda\lambda'}(\omega)$ or as a hole with amplitude $r_{\chi h}^{\lambda\lambda'}(\omega)$ into edge $\lambda'$. Likewise, transmission is possible with amplitude $t_{\chi e}^{\lambda\lambda'}(\omega)$ and $t_{\chi h}^{\lambda\lambda'}(\omega)$.
\subsection{Scattering state Green function}
From the scattering states, it is possible to construct the Green function of the system \cite{WMcMillan68, BLu2015}. With the equation of motion, it can be demonstrated in general, that for any time independent Hamiltonian and any $x\neq x'$, any Green function
\begin{equation}
\label{Eq:Green0}
\hat{G}^{R/A}(x,x',\omega)=\int\!dt e^{i(\omega \pm i 0^+) t}\hat{G}^{R/A}(x,x',t)
\end{equation}
with $\hat{G}^{R}(x,x',t)\!=\!-i\theta(t)\langle\!\lbrace\Psi(x,t),\Psi^{\dagger}(x',0)\rbrace\!\rangle$ and $\hat{G}^{A}(x,x',t)=\hat{G}^R(x',x,-t)^{\dagger}$, can be constructed from eigenstates $\Phi(x,\omega)$ of the Hamiltonian $\mathcal{H}(x, \hat{p}_x)$ and eigenstates $\tilde{\Phi}(x',\omega)$ of its transposed $\mathcal{H}^T(x', -\hat{p}_{x'})$, i. e.
\begin{equation}
\label{Eq:Green1}
\hat{G}_{n,m}^{R/A}(x,x',\omega)=\sum_{i,j}a_{i,j}^{R/A}(\omega)\phi_{i,n}(x,\omega)\tilde{\phi}_{j,m}(x',\omega),
\end{equation}
where $\phi_{i,n}$, $\tilde{\phi}_{j,m}$ are the $n$-th, $m$-th, component of the $i$-th, $j$-th eigenstate. The indices $i,j$ then run over all independent scattering states. 
Likewise, by integration of the equation $\left[\omega-\mathcal{H}(x,\hat{p}_x)\right]\hat{G}^{R/A}(x,x',\omega)=\delta(x-x')$, $\hat{G}^{R/A}(x,x',\omega)$ has to satisfy a discontinuity at $x=x'$
\begin{equation}
\label{Eq:Green2}
\lim_{\epsilon\rightarrow 0}\left[\hat{G}^{R/A}(x'+\epsilon,x',\omega)-\hat{G}^{R/A}(x'-\epsilon,x',\omega)\right]=C.
\end{equation}
For the Hamiltonian, given in Eq. (\ref{Eq:HamDens}), $C=i s_z\tau_z\sigma_z$.
Any function of the form of Eq. (\ref{Eq:Green1}), together with the constraint (\ref{Eq:Green2}), provides a valid Green function. However, for deriving a particular one, such as retarded or advanced, more information is needed to determine the coefficients $a_{i,j}^{R/A}(\omega)$. As all $a_{i,j}^{R/A}(\omega)$ are independent of $x$ and $x'$, it is sufficient to know the exact form of the Green function of interest in a single set of points $x_0$ and $x_0'$. Since it is significantly easier to calculate the Green function in a semi-infinite domain, it is suitable to position $x_0$ and $x_0'$ far in the left or right reservoirs.

We are interested in the superconducting pairing amplitude, that is related to the anomalous part of the retarded Green function. To compute the retarded Green function in the semi-infinite lead, we apply \textit{outgoing wave boundary conditions} \cite{SKash2000, WJHerrera2010}. We split the Green function into two parts $\hat{G}^R(x,x',\omega)=\hat{G}^R_<(x,x',\omega)\theta(x'-x)+\hat{G}^R_>(x,x',\omega)\theta(x-x')$. We explicitly derive the retarded Green function in the leftmost lead ($x_0~,x_0'<0$) in App. B. Then, we can construct the Green function of any $x$ and $x'$ as
\begin{equation}
\label{Eq:Green3}
\hat{G}^R(x,x',\omega)=U(x,x_0)\hat{G}^R(x_0,x_0',\omega)\tilde{U}^T(x',x_0'),
\end{equation} 
where $\tilde{U}(x',x_0')$ is the corresponding propagator derived from $\mathcal{H}^T(x',-\hat{p}_{x'})$. For our system, the Green function is a $8\times 8$ matrix with the general structure
\begin{equation}
\label{Eq:GreenFunctQPC}
\hat{G}^R(x,x',\omega)=\begin{pmatrix}
\hat{G}_{++}^R(x,x',\omega) & \hat{G}_{-+}^R(x,x',\omega)\\
\hat{G}_{+-}^R(x,x',\omega) & \hat{G}_{--}^R(x,x',\omega)\\
\end{pmatrix}.
\end{equation}
Each $\hat{G}_{\lambda\lambda'}^R(x,x',\omega)$ is itself a $4\times 4$ matrix, representing the intra edge Green function for $\lambda=\lambda'$ and inter edge Green function for $\lambda=-\lambda'$, respectively. Furthermore, each $\hat{G}_{\lambda\lambda'}^R(x,x',\omega)$ can be decomposed into
\begin{equation}
\label{Eq:Green4}
\hat{G}^R_{\lambda\lambda'}(x,x',\omega)=\begin{pmatrix}
\hat{G}^R_{\lambda\lambda',ee}(x,x',\omega) & \hat{G}^R_{\lambda\lambda',eh}(x,x',\omega)\\
\hat{G}^R_{\lambda\lambda',he}(x,x',\omega) & \hat{G}^R_{\lambda\lambda',hh}(x,x',\omega)\\
\end{pmatrix}.
\end{equation}
The off-diagonal parts of $G_{\lambda\lambda'}^R(x,x',\omega)$, thereby, carry the information about the superconducting pairing. In the basis of Eq. (\ref{Eq:Basis}), we can directly illustrate the spin-texture of the pairing with the decomposition into Pauli matrices
\begin{equation}
\label{Eq:Green4a}
\hat{G}^R_{\lambda\lambda',eh}(x,x',\omega)=f^R_{\lambda\lambda',0}(x,x',\omega)\sigma_0+f^R_{\lambda\lambda',j}(x,x',\omega)\sigma_j,
\end{equation}
where $j\in\{x,y,z\}$. In Eq. (\ref{Eq:Green4a}), $f^R_{\lambda\lambda',0}(x,x',\omega)$ is the singlet (S) component of the pairing, relating to the antisymmetric spin configuration $(\uparrow\downarrow\!-\!\downarrow\uparrow)$. Likewise, the triplet (T) components relate to the symmetric spin configuration with $f^R_{\lambda\lambda',z}(x,x',\omega)$ with $(\uparrow\downarrow\!+\!\downarrow\uparrow)$, and the equal-spin pairing $f^R_{\lambda\lambda',\uparrow\uparrow}=f^R_{\lambda\lambda',x}(x,x',\omega)- i f^R_{\lambda\lambda',y}(x,x',\omega)$, $f^R_{\lambda\lambda',\downarrow\downarrow}=f^R_{\lambda\lambda',x}(x,x',\omega)+ i f^R_{\lambda\lambda',y}(x,x',\omega)$, having $\uparrow\uparrow$, $\downarrow\downarrow$ configuration, respectively. From the definition of the advanced Green function $\hat{G}^A(x,x',\omega)=\hat{G}^R(x',x,\omega)^{\dagger}$, using Eq. (\ref{Eq:Green0}), we can translate the antisymmetry of the pairing amplitudes under exchange of the constituents into relations between retarded and advanced pairing amplitudes
\begin{eqnarray}
\label{Eq:Green5a}
f^R_{\lambda\lambda',0}(x,x',\omega)&=&f^A_{\lambda'\lambda,0}(x',x,-\omega),\\
\label{Eq:Green5b}
f^R_{\lambda\lambda',j}(x,x',\omega)&=&- f^A_{\lambda'\lambda,j}(x',x,-\omega)
\end{eqnarray}
with $j\in\{x,y,z\}$. Here, $f^A_{\lambda'\lambda,l}(x',x,\omega)$ ($l\in\{0,x,y,z\}$) is built from the advanced Green function.
Beside spin, orbit, and frequency, in case $\lambda\neq\lambda'$, we additionally have the edge index as degree of freedom to fulfill Eqs. (\ref{Eq:Green5a}) and (\ref{Eq:Green5b}) \cite{Tanaka2016b}. However, in this article, we will focus on the symmetry classification for one edge, i. e. $\lambda=\lambda'$. 

We can further decompose the symmetry requirements of Eqs. (\ref{Eq:Green5a}) and (\ref{Eq:Green5b}) into orbital and frequency symmetries. The orbital symmetries are captured by
\begin{equation}
\label{Eq:Green6a}
f_{\lambda\lambda,l}^R(x,x',\omega)=\zeta_{\lambda\lambda,l}^{R,+}(x,x',\omega)+\zeta_{\lambda\lambda,l}^{R,-}(x,x',\omega)
\end{equation}
with even (E) and odd (O) parts $\zeta_{\lambda\lambda,l}^{R,\pm}(x,x',\omega)=1/2(f^R_{\lambda\lambda,l}(x,x',\omega) \pm f^R_{\lambda\lambda,l}(x',x,\omega))$. To obey Eqs. (\ref{Eq:Green5a}) and (\ref{Eq:Green5b}), the symmetries in $\omega$ are then required to be
\begin{eqnarray}
\label{Eq:Green6b}
\zeta_{\lambda\lambda,0}^{R,\pm}(x,x',\omega)&=&\pm\zeta_{\lambda\lambda,0}^{A,\pm}(x,x',-\omega),\\
\label{Eq:Green6c}
\zeta_{\lambda\lambda,j}^{R,\pm}(x,x',\omega)&=&\mp\zeta_{\lambda\lambda,j}^{A,\pm}(x,x',-\omega),
\end{eqnarray}
where $j\in \{x,y,z\}$ and $\zeta_{\lambda\lambda,j}^{A,\pm}(x,x',-\omega),~\zeta_{\lambda\lambda,0}^{A,\pm}(x,x',-\omega)$ are the even and odd orbital parts of the advanced pairing amplitude. 
Eqs. (\ref{Eq:Green5a})-(\ref{Eq:Green6c}) describe all possible symmetries classes, that are, coined in the order frequency, spin, orbit as: ESE ($\zeta_{\lambda\lambda,0}^{R,+}(x,x',\omega)$), OSO ($\zeta_{\lambda\lambda,0}^{R,-}(x,x',\omega)$), ETO ($\zeta_{\lambda\lambda,j}^{R,-}(x,x',\omega)$) and OTE ($\zeta_{\lambda\lambda,j}^{R,+}(x,x',\omega)$).
\section{Local and non-local pairing symmetries}
As translation symmetry is broken in heterostructures, it is inevitable that any pairing is constituted by a mixture of even and odd orbital parts. Furthermore, since spin-momentum locking naturally implies triplet pairing, by construction ETO and OTE pairings are expected. In bare TI-SC heterojunctions, triplet pairing exists only in the form of the amplitude $f^R_{\lambda\lambda,z}(x,x',\omega)$ corresponding to the spin configuration $(\uparrow\downarrow\!+\!\downarrow\uparrow)$. It is, thus, difficult to discriminate this triplet amplitude in any (spin sensitive) conductance measurement from their singlet counterparts. This dilemma can be overcome if equal-spin pairing is generated in the heterojunction. 
Then, the CAR process across the junction, that is usually suppressed by spin-momentum locking \cite{PAdroguer2010}, is directly related to the injection of a Cooper pair with $\uparrow\uparrow$ ($\downarrow\downarrow$) spin-texture. 
A way to design $\uparrow\uparrow$ or $\downarrow\downarrow$ pairing at the helical edge is to include ferromagnetic ordering \cite{Crepin2015}. As it seems to be very difficult to combine ferromagnetic insulators and TIs in the laboratory, we identify a setup, in the absence of ferromagnetic ordering, in which (odd-frequency) $\uparrow\uparrow$- and $\downarrow\downarrow$-pairing at the helical edge can be generated. This is possible, due to the simultaneous presence of two TR invariant coupling terms, see Eqs. (\ref{Eq:QPC1}) and (\ref{Eq:QPC2}) above. By introducing a spin preserving coupling between the edges, Eq. (\ref{Eq:QPC1}) generates Andreev bound states between SC and QPC, that extend over both edges. Additionally, Eq. (\ref{Eq:QPC2}) breaks axial spin symmetry and, thus, allows for $\uparrow\uparrow$- and $\downarrow\downarrow$-pairing in each edge. 

\begin{figure}
\centering
\includegraphics[scale=0.18]{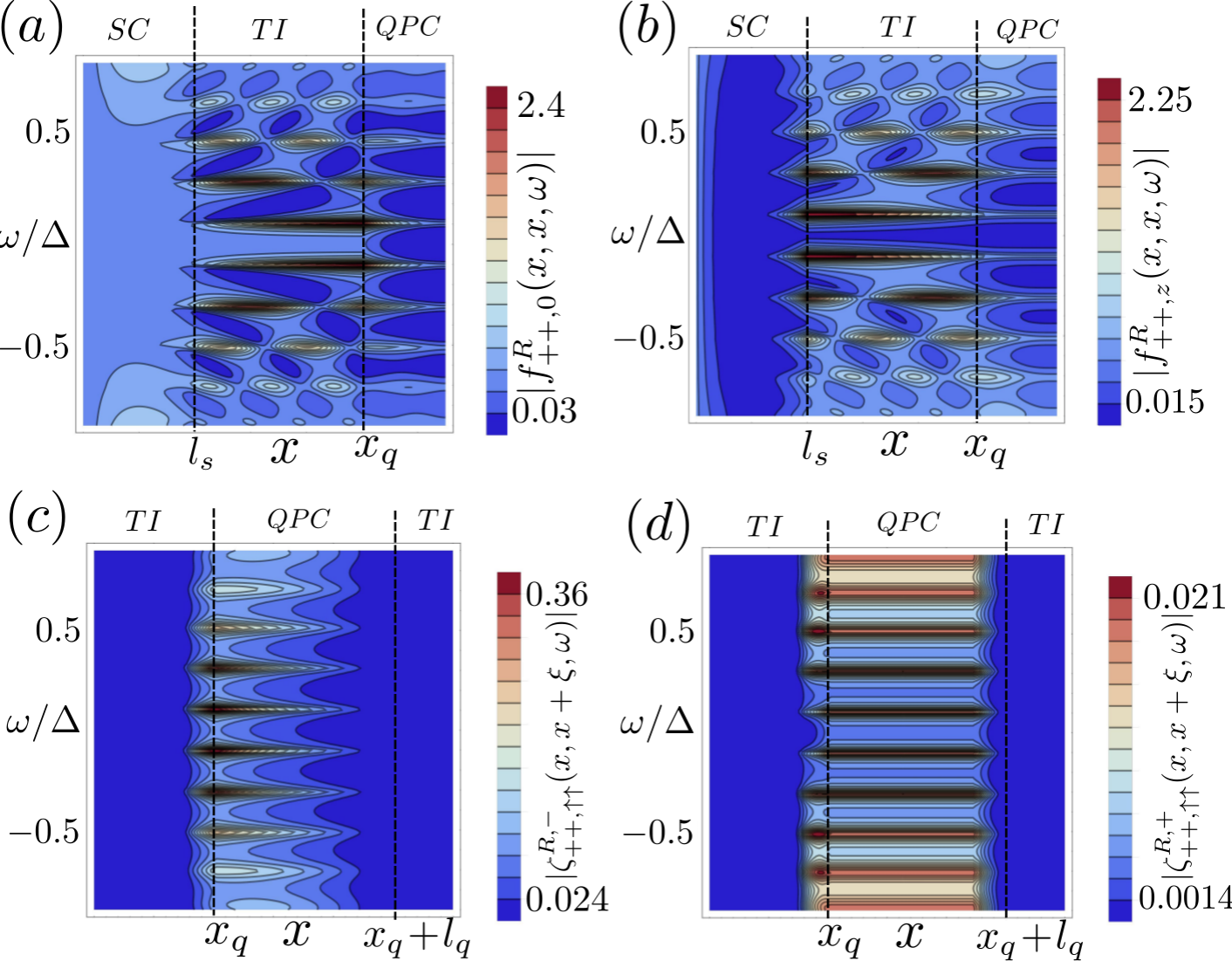}
\caption{Superconducting pairing in the heterojunction of Fig. \ref{Fig:sys} as a function of position $x$ and energy $\omega$. (a) and (b) illustrate the local singlet and triplet ($\uparrow\downarrow+\downarrow\uparrow$) pairing, while (c) and (d) show the even- and odd-frequency equal-spin pairing with $x'=x+\xi$ and $\xi=0.5\xi_{\Delta}$. We use the parameters: $\mu=0$, $t_0/\Delta=t_c/\Delta=t/\Delta=0.4$, $l_s=4 \xi_{\Delta}$, $l_q=3 \xi_{\Delta}$ and $x_q=10 \xi_{\Delta}$.}
\label{Fig:Results1}
\end{figure}

To demonstrate this effect, we proceed with the calculation of the pairing amplitudes. We apply Eq. (\ref{Eq:Green3}), with $G^R(x_0,x_0',\omega)$ for $x_0,~x_0'\rightarrow 0^-$ and calculate the amplitudes $f^R_{++,j}(x,x,\omega)$ with $j\in\{0,~z\}$ and $x\in[0,x_q+l_q]$. As $x=x'$, we naturally compute the ESE and OTE symmetries by construction. We find that local equal-spin correlations are totally suppressed throughout the whole junction $f^R_{++,\uparrow\uparrow}(x,x,\omega)=0$. However, non-local equal-spin correlations are present in the form of $f^R_{++,\uparrow\uparrow}(x,x+\xi,\omega)$ with $x\in[0,x_q+l_q-\xi]$. The results are depicted in Fig. \ref{Fig:Results1}. 

\begin{figure}
\centering
\includegraphics[scale=0.45]{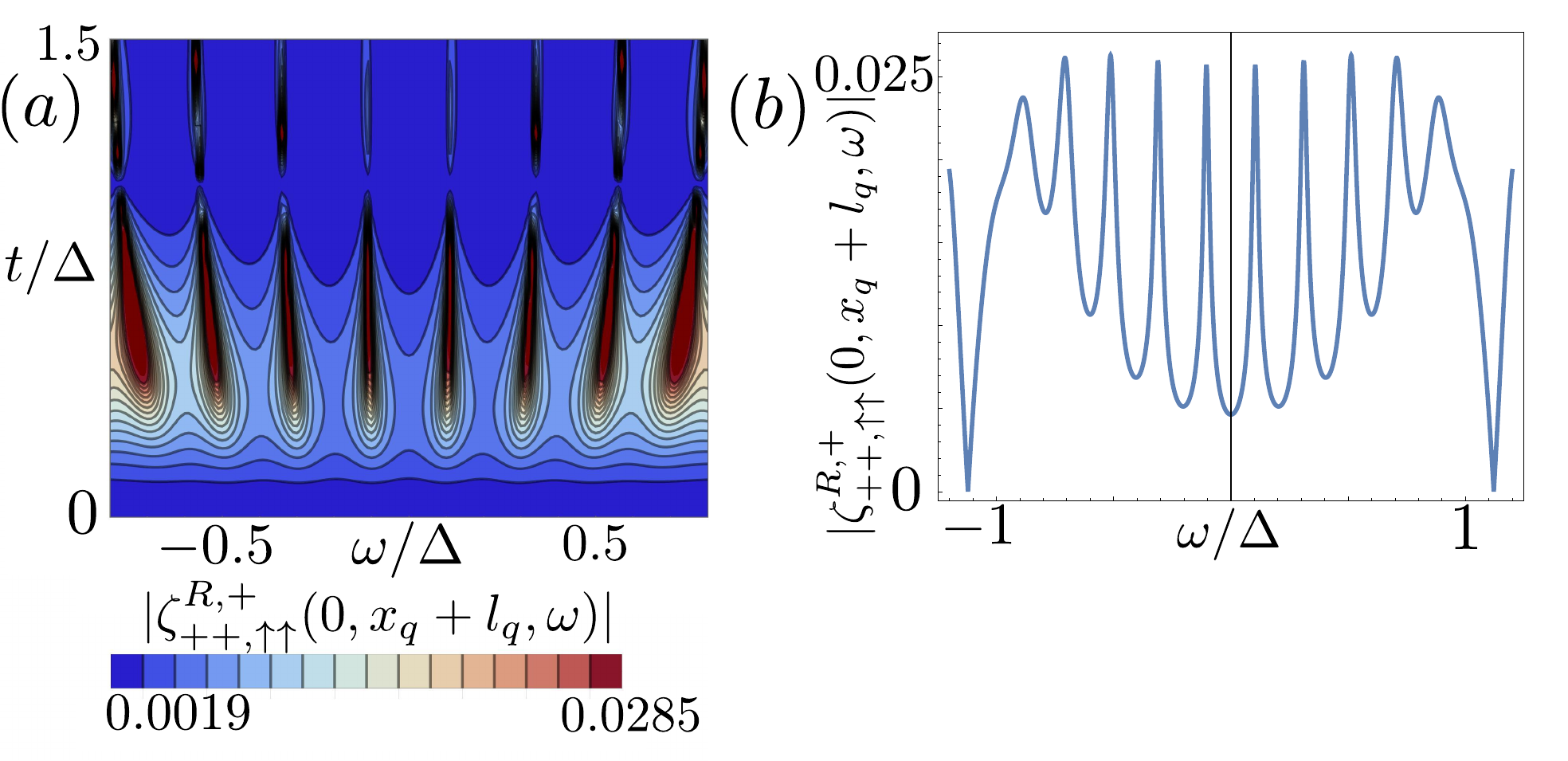}
\caption{(a) OTE part of the pairing amplitude $\vert f^R_{++,\uparrow\uparrow}(0,x_q+l_q,\omega)\vert$ as a function of coupling strength $t/\Delta=t_0/\Delta=t_c/\Delta$ and energy $\omega/\Delta$. (b) Line cut from (a) for $t/\Delta=t_0/\Delta=t_c/\Delta=0.4$ and the other parameters given in Fig. \ref{Fig:Results1}.}
\label{Fig:NonLocal}
\end{figure}

\begin{figure}
\centering
\includegraphics[scale=0.45]{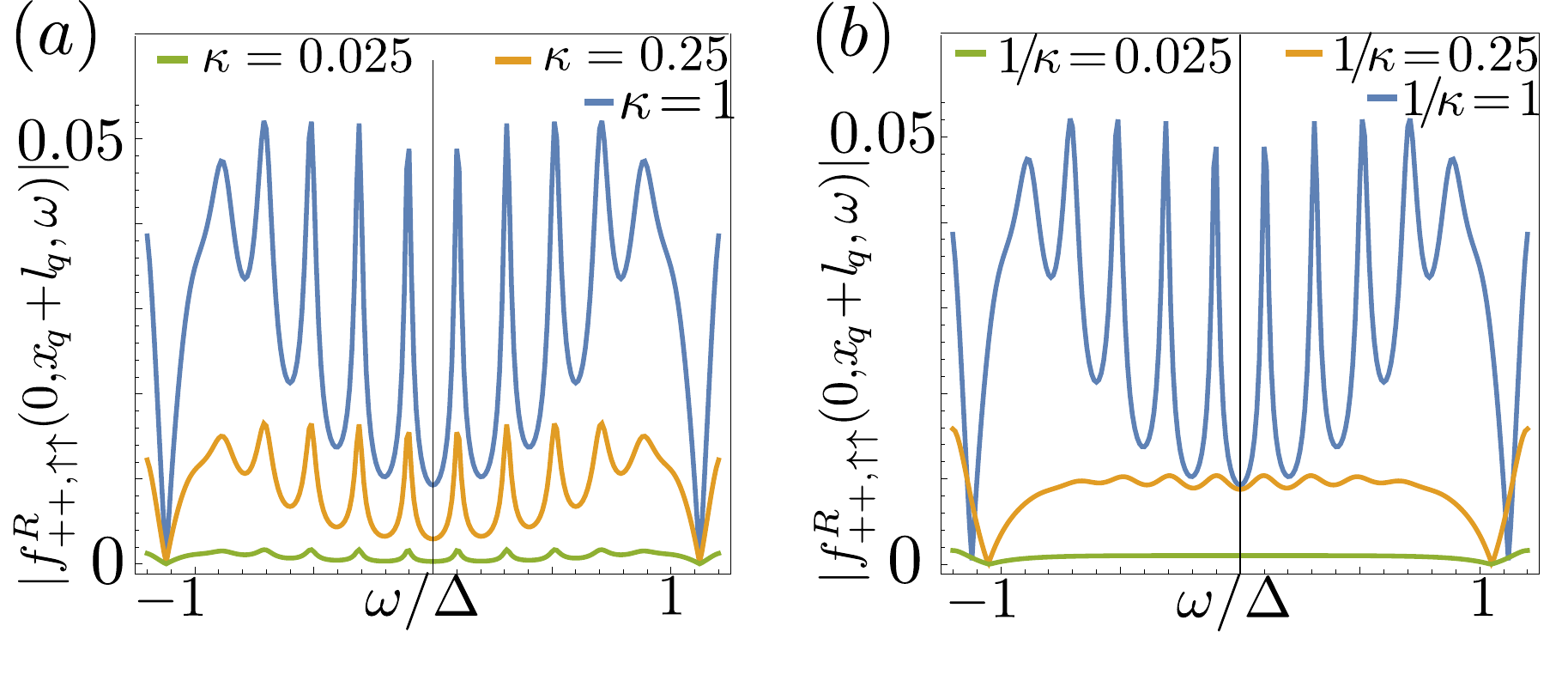}
\caption{Non-local $\uparrow\uparrow$-pairing as a function of $\omega$ for different values of the QPC coupling. We introduce the dimensionless parameter ratio $\kappa=t_c/t_0$. Other parameters are the same as in Fig. \ref{Fig:Results1}.}
\label{Fig:Results2}
\end{figure}

The suppression of local equal-spin pairing amplitudes is different with respect to the TR symmetry breaking case, where it is typically related to the amplitude of electron-electron (hole-hole) reflection $r_{ee}^{++}(\omega)$ ($r_{hh}^{++}(\omega)$), that vanish by TR symmetry in our system. The non-local equal-spin pairing $f^R_{++,\uparrow\uparrow}(x,x+\xi,\omega)$ created by the QPC, however, is finite whenever there is at least one point $\chi$ with $\chi\in[x,x+\xi]$ that belongs to the region of the QPC. This is realized, when at least one of the two spatial coordinates of the corresponding correlation function is part of the QPC region, i. e.  $x,x+\xi\in[x_q,x_q+l_q]$ (see Fig. \ref{Fig:Results1} (c) and (d)), or when the pairing happens across the whole QPC/junction (Fig. \ref{Fig:NonLocal}). In both cases, even- and odd-frequency parts appear (Fig. \ref{Fig:Results1} (c) and (d)). As expected, the non-local pairing $f^R_{++,\uparrow\uparrow}(0,x_q+l_q,\omega)$ has a maximum, whenever the energy of an Andreev bound state is matched (see. Fig. \ref{Fig:NonLocal}). Using Eqs. (\ref{Eq:Green6a}) and (\ref{Eq:Green6b}), we derive that the non-local equal-spin pairing across the junction is equally distributed from OTE and ETO parts. This turns out to be a very generic result, implied by spin-momentum locking. As the retarded Green function implements time-ordering, non-local equal-spin pairing from $x=0$ to $x'=x_q+l_q$ represents a correlation acting, for instance, forward in time and forward in space, while correlations from $x=x_q+l_q$ to $x'=0$ describe the corresponding process backward in space. For a defined pairing amplitude ($f^R_{\lambda\lambda,\uparrow\uparrow}(0,x_q+l_q,\omega)$ or $f^R_{\lambda\lambda,\downarrow\downarrow}(0,x_q+l_q,\omega)$) at the helical edge, only one of the two processes is finite due to spin-momentum locking. This behavior inverts as $x$ and $x'$ are exchanged. Consequently, from Eq. (\ref{Eq:Green6a}), we conclude that ETO and OTE pairing amplitudes are necessarily equal. Importantly, the presence of both coupling terms of Eqs. (\ref{Eq:QPC1}) and (\ref{Eq:QPC2}) is crucial for finite equal-spin pairing at a single edge (Fig. \ref{Fig:Results2}).

\begin{figure}
\centering
\includegraphics[scale=0.55]{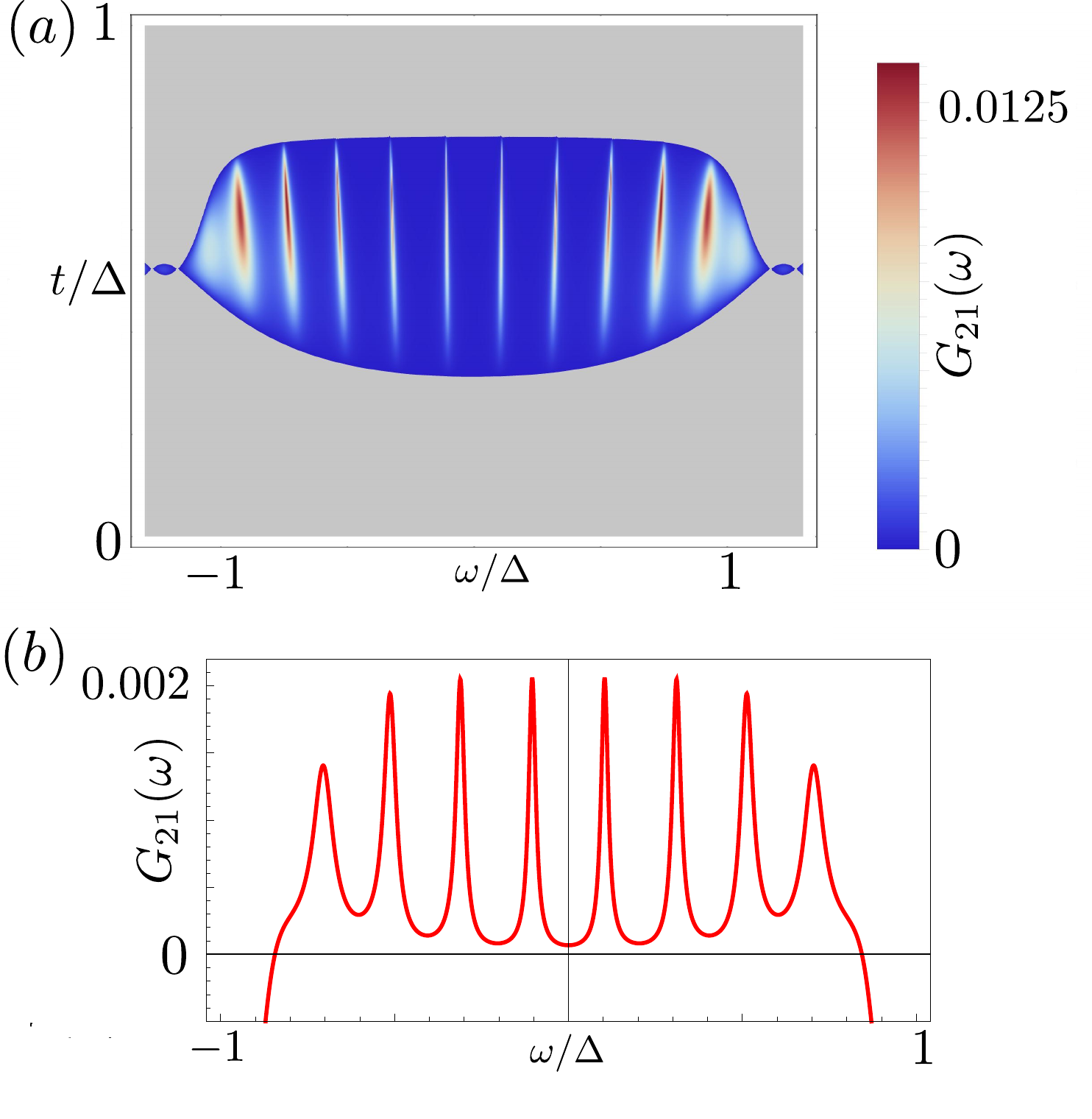}
\caption{(a) Non-local conductance $G_{21}(\omega)$ in units of $e^2/h$ for the setup shown in Fig. \ref{Fig:sys} with the parameters of Fig. \ref{Fig:Results1}. In the gray area we obtain $G_{21}(\omega)<0$, while in the colored region we have $G_{21}(\omega)>0$. (b) Line cut for $t/\Delta=0.4$.}
\label{Fig:cond}
\end{figure}

\section{Transport signatures}
As a consequence of spin-momentum locking in our structure, the observable directly related to the non-local equal-spin pairing is the CAR process. However, an incident particle, while transmitted, can either undergo CAR or EC. Since the two processes carry opposite charge, they enter with different signs in the non-local conductance. For concreteness, we are interested in the non-local differential conductance $G_{21}(\omega)$, measuring the transmission between contacts $1$ and $2$ of Fig. \ref{Fig:sys} at excitation energy $\omega$. By definition $G_{21}(\omega)$ is given as the differential change of the current at contact 2, when a voltage $V_1$ is applied at contact 1, i. e.
\begin{equation}
G_{21}(eV_1)=-\frac{\partial I_2(eV_1)}{\partial V_1}=\frac{e^2}{h}\left(\vert t_{eh}^{++}(eV_1)\vert^2-\vert t_{ee}^{++}(eV_1)\vert^2\right).
\end{equation}
The coefficients $t_{eh}^{++}(\omega)$ and $t_{ee}^{++}(\omega)$ are obtained from the corresponding scattering problem, defined in App. A. A clear evidence of CAR is, thus, only provided if $G_{21}(eV_1)>0$. {Unfortunately, this is not the generic case \cite{Morten2006, NM2003}, especially not at the helical edge, where CAR is additionally assigned to a spin configuration of the attributed Cooper pair \cite{Crepin2015}. Our system, however, has two major advantages in this respect: First, it provides non-local odd-frequency equal-spin pairing across the junction, directly related to $t_{eh}^{++}(\omega)$, without any TR breaking term. Second, as the QPC includes the other edge and we break axial spin symmetry, there are now three open channels available for EC. This possibility yields a reduced rate $t_{ee}^{\lambda\lambda'}(\omega)$ for each individual channel. The combination of those two effects leads to a domination of CAR over EC in a large domain of parameter space, which hence obeys a non-local conductance $G_{21}(\omega)>0$ (see. Fig. \ref{Fig:cond}). Notably, unlike other proposals, ours implies that nearly no finetuning is needed in order to measure CAR. Comparing Figs. \ref{Fig:NonLocal} and \ref{Fig:cond} (a), we notice that the area of positive non-local conductance is indeed related to the area of pronounced non-local equal-spin pairing, where in both cases we recognize the presence of Andreev bound states in the form of maxima.
\subsection{Influence of the chemical potential}
\begin{figure}
\centering
\includegraphics[scale=0.37]{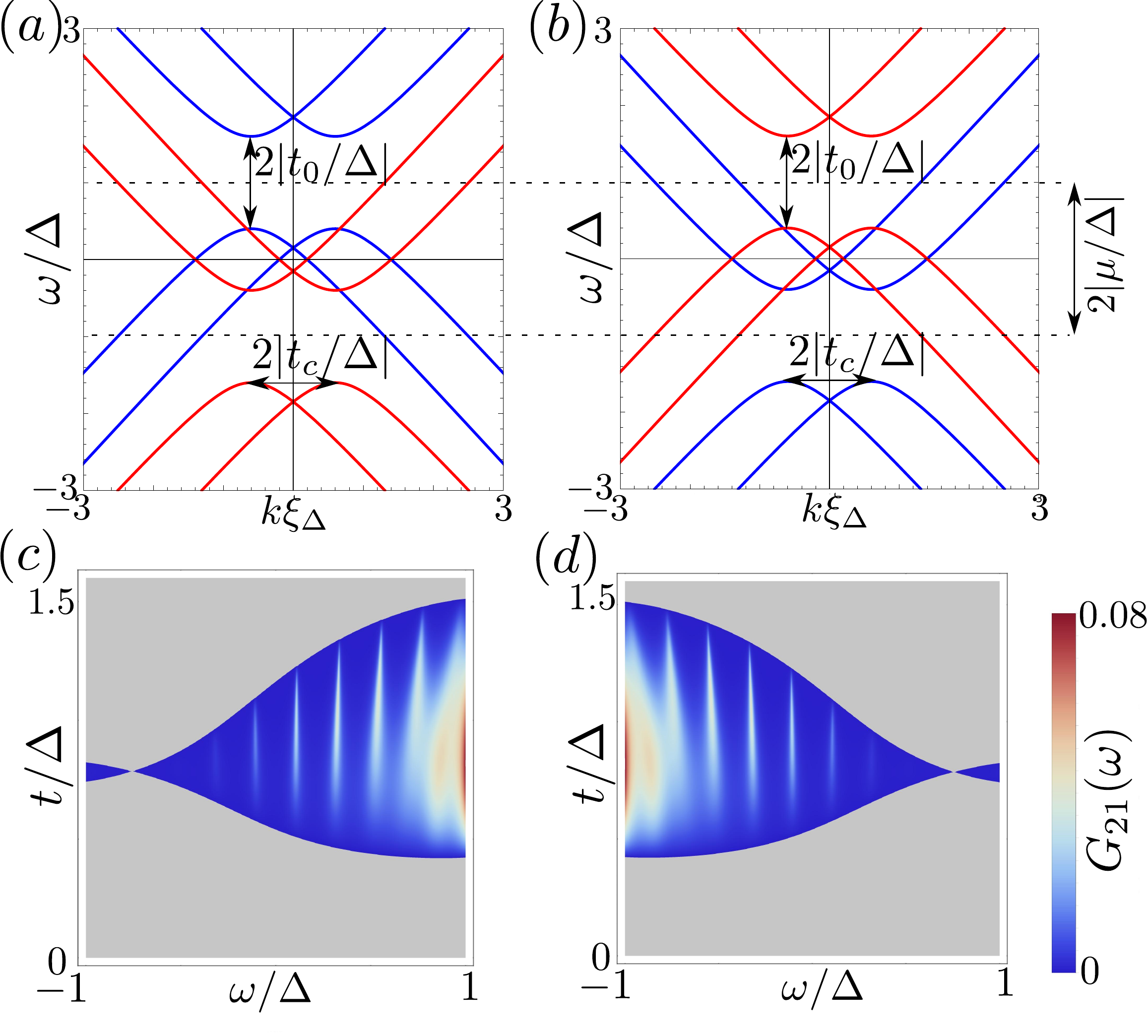}
\caption{(a)-(b) Dispersion relation of the QPC with the parameters $t_0/\Delta=t_c/\Delta=0.6$, and $\mu/\Delta=-1$ in (a) and $\mu/\Delta=1$ in (b). Electron-like excitations are shown in blue, hole-like excitations in red. (c)-(d) Non-local conductance $G_{21}(\omega)$ in units of $e^2/h$ for the parameters $l_s=3\xi_{\Delta}$, $l_q=2\xi_{\Delta}$, $x_q=9\xi_{\Delta}$ and $\mu/\Delta=-1$ in (c), $\mu/\Delta=1$ in (d), respectively.}
\label{Fig:disp}
\end{figure}
When including a finite chemical potential $\mu$ in the heterojunction, we obtain an asymmetry of the non-local conductance signature $G_{21}(\omega)$ with respect to $\omega\rightarrow -\omega$ (see Fig. \ref{Fig:disp} (c) and (d)). For positive chemical potential, negative excitation energies are favored in showing a non-local conductance $G_{21}(\omega)>0$ and vice versa. The explanation for this effect is found in the dispersion relation of the (infinitely extended) QPC (see Fig. \ref{Fig:disp} (a) and (b)). While the forward scattering $H_{t_c}$ is not affected by $\mu$, the backscattering term $H_{t_0}$ is sensitive to the chemical potential, which acts in opposite way to electron- and hole-like excitations. For $\mu\neq 0$ we can thus emphasize or suppress backscattering on electron- and hole-like states by finite excitation energies $\omega\neq 0$. To obtain $G_{21}(\omega)>0$ it is important to enhance the transmission of hole-like excitations and likewise suppress transmission of electron-like excitations across the QPC. Hence, comparing with Fig. \ref{Fig:disp} (a) and (b), for positive chemical potential, negative excitation energies are favourable and vice versa. This is indeed consistent with our results (see Fig. \ref{Fig:disp} (c)-(d)).
\section{Conclusion}
In this article, we have demonstrated the emergence of equal-spin triplet superconductivity at the helical edge without the need of ferromagnetic ordering. This can be achieved by combination of proximity-induced s-wave pairing and scattering off a quantum point contact (QPC). In the absence of axial spin symmetry, the QPC provides two possible coupling terms. In the presence of both, equal-spin triplet pairing is generated. While its local pairing amplitude is suppressed throughout the junction, non-local correlations arise whenever the spatial coordinates of the correlation function (partially) include the QPC or extend across it. On the basis of a symmetry analysis, we verify the (partial) odd-frequency nature of the non-local equal-spin pairing amplitude. This correlation is intimately related to the creation of equal-spin triplet Cooper pairs in the junction and, thus, to the process of crossed Andreev reflection (CAR). Notably, the QPC provides us with a direct access to this transmission channel, as it likewise lowers the rate of electron cotunneling (EC). Thus, the non-local conductance $G_{21}(\omega)$, given by the difference between CAR and EC, exhibits a positive signal. The domination of CAR over EC is given for a wide range in parameter space and, thus, persists without finetuning. The presence of CAR provides an unambignous evidence of odd-frequency superconductivity at the helical edge.

\begin{acknowledgments} We thank Felix Keidel for interesting discussions. We acknowledge financial support by the DFG (SPP1666 and SFB1170 "ToCoTronics"), the Helmholtz Foundation (VITI), the ENB Graduate school on "Topological Insulators", and the Studienstiftung des Deutschen Volkes.

\end{acknowledgments}

\appendix
\section{Scattering problem}
The scattering problem, defined by Eq. (\ref{Eq:scatter}), has eight independent solutions, classified according to the incident particle. Four of them, $\Phi_{in,1-4}(x)$, represent a particle incident from the left. At $x=0$ we have
\begin{eqnarray}
\Phi_{in,1}(0)&=&\left(1,r_{ee}^{++},r_{eh}^{++},0,r_{ee}^{+-},0,0,r_{eh}^{+-}\right)^T,\\
\Phi_{in,2}(0)&=&\left(0,r_{he}^{++},r_{hh}^{++},1,r_{he}^{+-},0,0,r_{hh}^{+-}\right)^T,\\
\Phi_{in,3}(0)&=&\left(0,r_{ee}^{-+},r_{eh}^{-+},0,r_{ee}^{--},1,0,r_{eh}^{--}\right)^T,\\
\Phi_{in,4}(0)&=&\left(0,r_{he}^{-+},r_{hh}^{-+},0,r_{he}^{--},0,1,r_{hh}^{--}\right)^T.
\end{eqnarray}
At $x=x_q+l_q$, the corresponding outgoing modes are
\begin{eqnarray}
\Phi_{out,1}(x_q+l_q)\!&=&\!\left(t_{ee}^{++},0,0,t_{eh}^{++},0,t_{ee}^{+-},t_{eh}^{+-},0\right)^T\!,\\
\Phi_{out,2}(x_q+l_q)\!&=&\!\left(t_{he}^{++},0,0,t_{hh}^{++},0,t_{he}^{+-},t_{hh}^{+-},0\right)^T\!,\\
\Phi_{out,3}(x_q+l_q)\!&=&\!\left(t_{ee}^{-+},0,0,t_{eh}^{-+},0,t_{ee}^{--},t_{eh}^{--},0\right)^T\!,\\
\Phi_{out,4}(x_q+l_q)\!&=&\!\left(t_{he}^{-+},0,0,t_{hh}^{-+},0,t_{he}^{--},t_{hh}^{--},0\right)^T\!.
\end{eqnarray}
All scattering amplitudes are functions of the excitation energy $\omega$, where the explicit dependence is dropped here for simplicity. Furthermore, each transmission amplitude has to be multiplied by a phase factor containing the position,
energy and chemical potential. However, since all transport properties do not depend on phases of the transmission amplitudes and as they do not enter into the lead Green function, derived below, we absorb these phases in the amplitudes.

Another set of four independent scattering states is constituted from a particle incident from the right $\Phi_{out,5-8}(x)$
\begin{eqnarray}
\Phi_{out,5}(\!x_q\!+\!l_q)\!&=&\!\left(\rho_{ee}^{++}\!,\!1\!,0,\rho_{eh}^{++},0,\rho_{ee}^{+-},\rho_{eh}^{+-},0\right)^T\!,\\
\Phi_{out,6}(\!x_q\!+l_q\!)\!&=&\!\left(\rho_{he}^{++}\!,\!0,\!1,\!\rho_{hh}^{++},0\!,\rho_{he}^{+-},\rho_{hh}^{+-},0\right)^{\!T\!} ,\\
\Phi_{out,7}(\!x_q\!+\!l_q)\!&=&\!\left(\rho_{ee}^{-+}\!,\!0,\!0,\!\rho_{eh}^{-+},\!1,\rho_{ee}^{--},\rho_{eh}^{--},0\right)^{\!T\!},\\
\Phi_{out,8}(\!x_q\!+\!l_q)\!&=&\!\left(\rho_{he}^{-+}\!,\!0,\!0,\!\rho_{hh}^{-+},\!0,\rho_{he}^{--},\rho_{hh}^{--},1\right)^{\!T\!},
\end{eqnarray}
partially transmitted to the left
\begin{eqnarray}
\Phi_{in,5}(0)&=&\!\left(0,\tau_{ee}^{++},\tau_{eh}^{++},0,\tau_{ee}^{+-},0,0,\tau_{eh}^{+-}\right)^T\!,\\
\Phi_{in,6}(0)&=&\!\left(0,\tau_{he}^{++},\tau_{hh}^{++},0,\tau_{he}^{+-},0,0,\tau_{hh}^{+-}\right)^T\!,\\
\Phi_{in,7}(0)&=&\!\left(0,\tau_{ee}^{-+},\tau_{eh}^{-+},0,\tau_{ee}^{--},0,0,\tau_{eh}^{--}\right)^T\!,\\
\Phi_{in,8}(0)&=&\!\left(0,\tau_{he}^{-+},\tau_{hh}^{-+},0,\tau_{he}^{--},0,0,\tau_{hh}^{--}\right)^T\!.
\end{eqnarray}
Each scattering problem of the form of Eq. (\ref{Eq:scatter}), is therefore an $8\times8$ linear eigenvalue problem. The complexity of the propagators $U_t(x,x')$, though, requires a numerical treatment of the problem.
\section{Green function for semi-infinite section}
The retarded Green function in a semi-infinite section, attached to a scattering region, can be calculated using \textit{outgoing wave boundary conditions}. We construct now the retarded Green function in the left lead of our system, i. e. for $x,~x'<0$. Therefore, we separate the Green function into 
\begin{equation}
\label{Eq:B1}
\hat{G}^R(x,x',\omega)=\hat{G}^R_<(x,x',\omega)\theta(x'-x)+\hat{G}^R_>(x,x',\omega)\theta(x-x'),
\end{equation}
and choose
\begin{eqnarray}
\label{Eq:B2}
\hat{G}^R_<(x,x',\omega)\!&=&\!\Phi_{in,1}(x)A_1^T(x')+\Phi_{in,2}(x)A_2^T(x')\nonumber\\
&+&\!\Phi_{in,3}(x)A_3^T(x')\!+\!\Phi_{in,4}(x)A_4^T(x'),\\
\label{Eq:B3}
\hat{G}^R_>(x,x',\omega)\!&=&\!\Phi_{in,5}(x)A_5^T(x')+\Phi_{in,6}(x)A_6^T(x')\nonumber\\
&+&\Phi_{in,7}(x)A_7^T(x')\!+\!\Phi_{in,8}(x)A_8^T(x')
\end{eqnarray}
with the unknown vectors $A_j(x')^T=\left(A_{j1},A_{j2},A_{j3},A_{j4}\right)^T$. Furthermore, the eigenstates $\Phi_{in,j}(x)$ are given by
\begin{equation}
\Phi_{in,j}(x)=U_{0}(x,0)\Phi_{in,j}(0).
\end{equation}
Inserting Eqs. (\ref{Eq:B2}) and (\ref{Eq:B3}) in Eq. (\ref{Eq:Green2}), this yields the solution for the vectors $A_j(x')$ and, thus, the form of the retarded Green function in the TI leftmost of the scattering region for $x,x'<0$.
Adopting the notation of Eq. (\ref{Eq:GreenFunctQPC}), we obtain

\onecolumngrid
\begin{equation}
\label{Eq:B4}
\hat{G}^R_{++}(x,x',\omega)=-i\begin{pmatrix}
e^{i(x-x')(\mu+\omega)}\theta(x-x') & 0 & 0 & 0 \\
e^{-i(x+x')(\mu+\omega)}r_{ee}^{++} & e^{-i(x-x')(\mu+\omega)}\theta(x'-x) & 0 & e^{i(x'(\mu-\omega)-x(\mu+\omega))} r_{he}^{++} \\
e^{i(x(\mu-\omega)-x'(\mu+\omega))}r_{eh}^{++} & 0 & e^{i(x-x')(\mu-\omega)}\theta(x'-x) & e^{i(x+x')(\mu-\omega)}r_{hh}^{++}\\
0 & 0 & 0 & e^{-i(x-x')(\mu-\omega)}\theta(x-x')\\
\end{pmatrix},
\end{equation}
\begin{equation}
\label{Eq:B5}
\hat{G}^R_{--}(x,x',\omega)=-i\begin{pmatrix}
e^{-i(x-x')(\mu+\omega)}\theta(x'-x) & e^{-i(x+x')(\mu+\omega)}r_{ee}^{--} & e^{i(x'(\mu-\omega)-x(\mu+\omega))} r_{he}^{--} & 0 \\
0 & e^{i(x-x')(\mu+\omega)}\theta(x-x') & 0 & 0 \\
0 & 0 & e^{-i(x-x')(\mu-\omega)}\theta(x-x') & 0\\
0 & e^{i(x(\mu-\omega)-x'(\mu+\omega))}r_{eh}^{--} & e^{i(x+x')(\mu-\omega)}r_{hh}^{--} &
e^{i(x-x')(\mu-\omega)}\theta(x'-x)\\
\end{pmatrix},
\end{equation}
\begin{equation}
\label{Eq:Off1}
\hat{G}_{-+}^R(x,x',\omega)=-i\begin{pmatrix}
0 & 0 & 0 & 0\\
0 & e^{-i(x+x')(\mu+\omega)}r_{ee}^{-+} & e^{i(x'(\mu-\omega)-x(\mu+\omega))}r_{he}^{-+} & 0 \\
0 & e^{i(x(\mu-\omega)-x'(\mu+\omega))}r_{eh}^{-+} & e^{i(x+x')(\mu-\omega)}r_{hh}^{-+} & 0 \\
0 & 0 & 0 & 0 \\
\end{pmatrix},
\end{equation}
and
\begin{equation}
\label{Eq:Off2}
\hat{G}_{+-}^R(x,x',\omega)=-i\begin{pmatrix}
e^{-i(x+x')(\mu+\omega)}r_{ee}^{+-} & 0 & 0 & e^{i(x'(\mu-\omega)-x(\mu+\omega))}r_{he}^{+-}\\
0 & 0 & 0 & 0\\
0 & 0 & 0 & 0\\
e^{i(x(\mu-\omega)-x'(\mu+\omega))}r_{eh}^{+-} & 0 & 0 & e^{i(x+x')(\mu-\omega)}r_{hh}^{+-}\\
\end{pmatrix}.
\end{equation}

\twocolumngrid
Eqs. (\ref{Eq:B4})-(\ref{Eq:Off2}) define the lead Green function that is used in the main text to construct the Green function anywhere in the heterojunction.

\twocolumngrid

\end{document}